\RequirePackage[T1]{fontenc}
\documentclass[12pt]{article}
\pdfoutput=1 % if your are submitting a pdflatex (i.e. if you have images in pdf, png or jpg format)
\usepackage[height=8.85in,width=6.45in]{geometry}

\usepackage[utf8]{inputenc}
\usepackage{amsmath}
\usepackage{amssymb}
\usepackage{mathtools}
\numberwithin{equation}{section}
\usepackage{slashed}
\usepackage{braket}
\usepackage[svgnames]{xcolor}
\usepackage[colorlinks,linktocpage=true,citecolor=DarkGreen,linkcolor=FireBrick]{hyperref}
\usepackage{cite}
\usepackage{graphicx}
\usepackage{tikz}
\usetikzlibrary{decorations.markings}
\usepackage{tikz-cd}
\usepackage{times}
\usepackage[scaled]{couriers}
\usepackage{bm}
\usepackage{subfig}
\usepackage{mathrsfs}
\usepackage{amsthm}

\usepackage{tikz}
\usetikzlibrary{arrows.meta}

\usepackage{xcolor}
\usepackage{mdframed}

\renewenvironment{figure}[1][]{
  \begin{originalfigure}[#1]
    \begin{mdframed}[linecolor=black!0,backgroundcolor=black!1]
}{
    \end{mdframed}
  \end{originalfigure}
}

\theoremstyle{plain}

\theoremstyle{definition}

\def\d{{\rm d}}
\def\i{{\mathsf i}}

\DeclareMathOperator{\tr}{tr}

\def\diag{\mathop{\rm diag}\nolimits}

\usepackage{slashed}

\def\cA{{\cal A}}

\def\cG{{\cal G}}

\def\cL{{\cal L}}

\def\cZ{{\cal Z}}

\def\bC{{\mathbb C}}

\def\bR{{\mathbb R}}

\def\bZ{{\mathbb Z}}

\def\sC{{\mathsf C}}

\def\sF{{\mathsf F}}

\def\sL{{\mathsf L}}
\def\sM{{\mathsf M}}

\def\sR{{\mathsf R}}

\def\sT{{\mathsf T}}

\def\sg{{\mathsf g}}

\def\so{{\mathsf o}}

\def\U{\mathrm{U}}
\def\SU{\mathrm{SU}}
\def\O{\mathrm{O}}
\def\SO{\mathrm{SO}}
\def\Sp{\mathrm{Sp}}

\def\Spin{\mathrm{Spin}}
\def\Pin{\mathrm{Pin}}

\def\so{\mathfrak{so}}

\def\g{\mathfrak{g}}

\def\beq#1\eeq{\begin{align}#1\end{align}}

\usepackage[all]{xy}

\begin{document}

\begin{titlepage}

\begin{flushright}
TU-1296
\end{flushright}

\vskip 3cm

\begin{center}

{\Large \bfseries The absence of global anomalies of CP symmetry  }

\vskip 1cm

Kazuya Yonekura~$^1$
\vskip 1cm

\begin{tabular}{ll}
$^1$ & Department of Physics, Tohoku University, Sendai 980-8578, Japan
\end{tabular}

\vskip 1cm

\end{center}

\noindent
Some solutions to the strong CP problem assume that CP symmetry is a gauge symmetry, which is then spontaneously broken. For this scenario to be possible, the CP symmetry should not have any nonperturbative (global) anomalies. In this paper, we study anomalies of CP symmetry of fermions which are coupled to gravity and gauge fields with a gauge group $G$. When $G$ is connected and simply connected, we show that gauging a CP symmetry does not produce any new anomaly beyond the one before gauging it. In particular, the standard model matter content does not have anomalies.

\end{titlepage}

\setcounter{tocdepth}{3}
%\tableofcontents

%\newpage

\tableofcontents

%\newpage

%%%%%%%%%%%%%%%%%%%%%%%%%%%%%%%%%%%%
\section{Introduction and summary}
%%%%%%%%%%%%%%%%%%%%%%%%%%%%%%%%%%%%

The standard model itself does not have an exact CP symmetry because of the CP-violating phase in the Yukawa couplings. However, the effective $\theta$ angle of QCD is experimentally constrained to be extremely small, suggesting an approximate CP symmetry in the QCD sector. This is the famous strong CP problem. The axion is one of the most popular solutions to the strong CP problem. Another possibility is that the CP symmetry is an exact symmetry of some UV theory, which is then spontaneously broken at low energies (see e.g. \cite{Nelson:1983zb,Barr:1984qx,Babu:1988mw,Babu:1989rb,Barr:1991qx,Lavoura:1997pq,Vecchi:2014hpa,Dine:2015jga,Hall:2018let,Dunsky:2019api,Choi:2019omm,Craig:2020bnv,Valenti:2021rdu,Fujikura:2022sot,Girmohanta:2022giy,Bonnefoy:2023afx,Nakagawa:2024ddd,Murai:2024alz,Hall:2024qqe,Feruglio:2024ytl,Murai:2024bjy,Hisano:2025tud,Liang:2025dkm,Bonnefoy:2025rvo,Csaki:2025ikr,Kobayashi:2025thd,Kobayashi:2025rpx,Benabou:2025viy}). This scenario is consistent due to the smallness of radiative corrections from the Yukawa couplings to the effective $\theta$ angle~\cite{Ellis:1978hq} (see also \cite{Hisano:2023izx,Banno:2023yrd,Banno:2025pfq} for recent developments). 

A CP symmetry is either a global symmetry or a gauge symmetry. Global symmetry is not a fundamental principle of particle physics. (It is just a convenient tool when an approximate global symmetry exists.) In fact, it is believed that there is no exact global symmetry in quantum gravity~\cite{Hawking:1975vcx,Banks:1988yz,Coleman:1988cy,Giddings:1988cx,Kallosh:1995hi,Arkani-Hamed:2006emk,Banks:2010zn,Harlow:2018tng}. Therefore, if the CP symmetry is not gauged, it is natural to expect that it would be explicitly broken in the UV, producing the $\theta$ term in QCD. Since the strong CP problem is a naturalness problem, it is more natural to consider gauged CP symmetry.

To consider gauged CP symmetry, we need to make sure that there is no anomaly. Before gauging the CP symmetry, it is a textbook fact that the usual perturbative gauge and gravitational anomalies (produced by triangle diagrams) are cancelled in the standard model. However, perturbative anomaly cancellation is not enough. Nonperturbative, or global, anomalies must also be cancelled. Perhaps the most famous global anomaly is the Witten $\mathrm{SU}(2)$ anomaly \cite{Witten:1982fp}, which is known to be absent in the standard model. In principle, there can be other global anomalies. Fortunately, it has been shown that there is no global anomaly in the standard model without gauged CP symmetry \cite{Freed:2006mx,Garcia-Etxebarria:2018ajm,Witten:2019bou}. 

The main purpose of this paper is to show that there is no new anomaly when the CP symmetry is gauged. We emphasize that this question is nontrivial. Indeed, in three spacetime dimensions, there exist very interesting global anomalies of (C)P symmetry which are relevant to condensed matter physics such as topological insulators and superconductors~\cite{Hsieh:2015xaa,Witten:2015aba}. 

More generally, the computations and applications of global anomalies in models beyond the standard model are interesting (see e.g. \cite{Hsieh:2018ifc,Garcia-Etxebarria:2018ajm,Davighi:2019rcd,Wang:2018jkc,Wan:2019gqr,Wang:2020xyo,Costa:2020krs,Wang:2020gqr,Smith:2021vbf,Wang:2021ayd,Davighi:2022fer,Davighi:2022icj,Wang:2022osr,Putrov:2023jqi,Kawasaki:2023mjm,Kawasaki:2023zpd,Wang:2025oow,Wan:2025lad,Wan:2024kaf}.) We will review how to think about global anomalies in a systematic approach, based on \cite{Witten:1999eg,Witten:2015aba,Witten:2019bou}.

\paragraph{The main result.} We consider a four dimensional theory with gauge group $G$. The $k$-th homotopy group of $G$ is denoted by $\pi_k(G)$. Then our main result may be summarized as follows:\footnote{Anomalies of CP symmetry in four dimensions are briefly mentioned in \cite{Garcia-Etxebarria:2018ajm}, but our discussions will be different from those of \cite{Garcia-Etxebarria:2018ajm} due to the details of the CP symmetry. As we explain in Section~\ref{sec:CP}, the relevant symmetry group for our purposes is a semidirect product $\mathrm{Pin}^+(d) \ltimes G$ rather than an ordinary product. For instance, $\Omega_5^{\mathrm{Pin}^+}(B \mathrm{SU}(2))=0$ but $\Omega_5^{\mathrm{Pin}^+ \ltimes \mathrm{SU}(2)}( \mathrm{pt}) \simeq \mathbb{Z}_2$ is nonzero, detected by the ordinary Witten $\mathrm{SU}(2)$ anomaly. See also Section~\ref{sec:perturbative} for the situation of perturbative anomalies.} 
\begin{itemize}
    \item Suppose that the gauge group $G$ is connected and simply connected, $\pi_0(G)=\pi_1(G)=0$. Suppose moreover that there is no anomaly before gauging a CP symmetry. Then, there is no new anomaly after gauging it.
\end{itemize}
From this result, the case of the standard model matter content can be understood as follows. The gauge group of the standard model can be embedded into grand unified gauge groups such as $G=\mathrm{SU}(5)$ and $\mathrm{Spin}(10)$ which satisfy the above conditions. The standard model matter content has no anomaly under these groups~\cite{Freed:2006mx,Garcia-Etxebarria:2018ajm,Witten:2019bou} before gauging the CP symmetry. Therefore, by our result, there is no new anomaly after gauging the CP symmetry.

Our method will be just a straightforward extension of the one in \cite{Witten:2019bou} which was used to study global anomalies of gauge groups with $\pi_0(G)=\pi_1(G)=0$ in four dimensions.

The rest of the paper is organized as follows. In Section~\ref{sec:CP}, we recall some elementary facts about CP symmetry, and review what it means to gauge it. In Section~\ref{sec:anomalies}, we study anomalies in the presence of CP symmetry. In Section~\ref{sec:string}, we briefly discuss string theory realization.

%%%%%%%%%%%%%%%%%%%%%%%%%%%%%%%%%%%%
\section{CP symmetry}\label{sec:CP}
%%%%%%%%%%%%%%%%%%%%%%%%%%%%%%%%%%%%
CP symmetry is a basic topic in quantum field theory. However,
to state the results of global anomalies associated to CP symmetries precisely, it is necessary to clarify some details of CP symmetries and their gauging. These issues have been discussed in \cite{Witten:2015aba}, particularly in the case of three dimensions motivated by topological insulators and superconductors. Here we discuss the case of four dimensions. Most of this section is elementary, but we will try to be careful, and we also review some mathematical concepts which might not be so well-known in particle phenomenology.  

We take the spacetime metric to have either Lorentzian, mostly plus signature $g_{\mu\nu}=\diag (-+++)$, or Euclidean signature $g_{\mu\nu}=\diag (++++)$, the latter being obtained after Wick rotation. The gamma matrices are chosen such that $\{\gamma_\mu, \gamma_\nu \}=2g_{\mu\nu}$. For a 4-component fermion $\Psi$ in Lorentzian signature, we define $\bar\Psi = \Psi^\dagger \beta$ as usual, where $\beta = \i  \gamma^0$. (In our convention, the spatial gamma matrices $\gamma^i~(i=1,2,3)$ are Hermitian and $\gamma^0$ is anti-Hermitian in Lorentzian signature.) In Euclidean signature, $\Psi$ and $\bar\Psi$ may be regarded as independent variables in Euclidean path integrals. 

\subsection{The CP symmetry of a single neutral Weyl fermion} \label{sec:singleneutral}

Let us first start from the simple case of a neutral Dirac fermion $\Psi$ with the Lagrangian given by
\beq
\cL = -\bar\Psi(\slashed{\partial}+m)\Psi, \label{eq:diraclagrangian}
\eeq
where $\slashed{\partial}=\gamma^\mu \partial_\mu$ and $m$ is the mass parameter.
In Minkowski spacetime, consider the transformation
\beq
x^\mu \to x'^\mu=(\delta^\mu_\nu -2n^\mu n_\nu)x^\nu. \label{eq:xtransf}
\eeq
where $n^\mu$ is a spacelike unit vector ($n^2=1$). This transformation is a reflection along the direction $n^\mu$. 
One can check that the Lagrangian is invariant under 
\beq
\sR_n: \Psi(x) \to \Psi'(x)= \alpha \gamma_5 \slashed{n} \Psi(x'), \qquad \bar\Psi(x) \to \bar\Psi'(x)= \alpha^*   \bar\Psi(x')  \slashed{n}  \gamma_5 \label{eq:parity1}
\eeq
where $\alpha \in \U(1)$ is an arbitrary phase, $\gamma_5$ is the usual chirality operator, and $\slashed{n}=n^\mu \gamma_\mu$. The appearance of $\slashed{n}$ is natural from Lorentz covariance, 
and we indeed have 
\beq
\slashed{n}\gamma_5 \gamma^\mu \gamma_5 \slashed{n} = (\delta^\mu_\nu - 2n^\mu n_\nu)\gamma^\nu,
\eeq
which is the same transformation as \eqref{eq:xtransf}. 
The presence of $\gamma_5$ in $\sR_n$ is necessary to make the mass term invariant. The transformation $\sR_n$ in \eqref{eq:parity1} is a parity (rather than CP) transformation, or more precisely a reflection transformation along the direction $n^\mu$.

Next, we consider a neutral Weyl fermion. Instead of using a 2-component spinor, let us impose a Majorana condition on $\Psi$. (Gauge charges will be discussed later). The Majorana condition is given by
\beq
\bar \Psi = \Psi^T C,
\eeq
where $C$ is a matrix which needs to satisfy the following conditions. The Lagrangian
\beq
\cL = -\frac{1}{2} \Psi^T( (C\gamma^\mu)\partial_\mu +m C)\Psi,
\eeq
which is obtained by replacing $\bar\Psi \to \Psi^T C$ in \eqref{eq:diraclagrangian} and dividing by $2$, must make sense. This requires that $ (C\gamma^\mu)\partial_\mu$ and $C$ are antisymmetric (since $\Psi$ is Grassmann), and hence $C$ is antisymmetric while $C\gamma^\mu $ is symmetric. Thus
\beq
C^T=-C, \qquad C^{-1}\gamma_\mu^T C = -\gamma_\mu, \qquad C^{-1}\gamma_5^T C = \gamma_5, \label{eq:conditiononC}
\eeq
where the last equation follows from $\gamma_5 \propto \gamma^0\gamma^1\gamma^2\gamma^3$.\footnote{More explicitly, if $\gamma_0^T=\gamma_0$, $\gamma_2^T=\gamma_2$, $\gamma_1^T=-\gamma_1$ and $\gamma_3^T=-\gamma_3$, we may take $C \propto \gamma^0 \gamma^2$.}

One can check that $\bar \Psi$ and $\Psi^T C$ transform in the same way under the usual Lorentz symmetry without a reflection. On the other hand, under the reflection $\Psi(x) \to \Psi'(x)= \alpha \gamma_5 \slashed{n} \Psi(x')$ we get
\beq
\sR_n : \Psi^T (x) C \to \Psi'^T(x) C = - \alpha \Psi^T(x') C \slashed{n} \gamma_5,
\eeq
where we have used \eqref{eq:conditiononC}. By comparing it with $\bar\Psi(x) \to \bar\Psi'(x)= \alpha^*  \bar\Psi(x')  \slashed{n}  \gamma_5$, we see that $\alpha$ must be pure imaginary, $\alpha = \pm \i$, for $\bar\Psi$ and $\Psi^T C$ to have the same transformation. In particular, $(\alpha \gamma_5 \slashed{n})^2=1$ and we get
\beq
\sR_n^2=1. \label{eq:pin+condition}
\eeq
For a neutral Majorana particle, there is no distinction between P and CP transformations (i.e., the charge conjugation is trivial). Therefore, \eqref{eq:pin+condition} is valid also for CP, or more precisely CR where R stands for reflection.

We have derived \eqref{eq:pin+condition} by a rather elementary computation, but there is a general reason for it when there is only a single Weyl fermion. Let $\sT$ be a time reversal symmetry. The combination $\sR_n \sT$ always exists, which is usually called the CPT theorem but may be more appropriately called the CRT theorem~\cite{Witten:2015aba}. %(In the current discussion of a single neutral Weyl fermion, the charge conjugation is trivial.)
For simplicity, we denote $\sR_n$ by just $\sR$. Let $\tau$ be the Euclidean time direction. We take the direction of $n^\mu$ to be orthogonal to the Euclidean time direction. In Euclidean space, $\sR \sT$ corresponds to just a $\pi$ rotation in the $\tau n$ plane. (This is the reason that CRT always exists in $d \geq 2$ dimensions: it is just a $\pi$ rotation in Euclidean signature.) On fermions, this $\pi$ rotation is realized by $\exp(\frac{\pi}{2} \gamma^\tau \gamma^n)= \gamma^\tau \gamma^n$, where $\gamma^\tau$ and $\gamma^n$ are the gamma matrices in the directions of $\tau$ and $n$, respectively (and hence in particular $\gamma^n=\slashed{n}$). It satisfies $( \gamma^\tau \gamma^n)^2=-1$. However, in Lorentzian signature, the time direction $t$ is Wick-rotated from $\tau$ such that $\gamma^t = -\i \gamma^\tau$ and hence $(\gamma^t \gamma^n)^2=1$. Essentially, this fact implies $(\sR \sT)^2=1$ in Lorentzian signature. This result is indeed known to be generally true in the CRT theorem. 

Let us also notice that $\sR \sT=(-1)^{\sF} \sT \sR$, where $(-1)^\sF$ is the fermion parity operator which acts on fermions as $(-1)^\sF=-1$. This is essentially because $\gamma^t\gamma^n = - \gamma^n \gamma^t$. 

Moreover, we have $\sT^2=(-1)^\sF$ for a single Weyl fermion in four dimensions. This can be understood from the fact that $\sT$ commutes with the spatial $\SU(2)$ rotational symmetry and is also antilinear. Then it is an antilinear transformation which maps pseudo-real (half integer spin) representations of $\SU(2)$ to themselves. From a general fact (or actually a definition) about pseudo-real representations, such an antilinear map squares to $-1$. Integer spin representations are strictly-real and hence $\sT^2$ acts as $+1$ on them. Therefore $\sT^2=(-1)^\sF$. 

Combining the above facts, we get
\beq
1=(\sR \sT)^2 = (-1)^\sF \sT^2 \sR^2 = \sR^2. \label{eq:pin+condition2}
\eeq
This is the same result as \eqref{eq:pin+condition}. What we have shown here is that \eqref{eq:pin+condition} is a general result for a single Weyl fermion.  

If the number of Weyl fermions is even, the result \eqref{eq:pin+condition} is not necessarily true. This is because we can construct a strictly-real representation from the direct sum of two copies of a pseudo-real representation.
For instance, if there are two fermions $(\Psi_1, \Psi_2)$, we can combine the above reflection with a transformation 
$(\Psi_1, \Psi_2) \to (\Psi_2, -\Psi_1)$
to realize $\sR^2=(-1)^\sF$. This situation arises naturally e.g., in the compactification of a five-dimensional theory on $S^1$ to four dimensions. We do not study such a case in the present paper. 

\subsection{The Pin groups}
For the study of global anomalies, it is necessary to specify symmetry groups precisely. First let us recall basic facts when there is no CP symmetry. 

Without fermions, the Lorentz symmetry group in $d$ dimensions can be taken to be $\SO(d-1,1)$, or $\SO(d)$ after Wick rotation.
However, in the presence of fermions, it is not possible to take $\SO(d)$ as the Lorentz symmetry group. The minimal extension in this case is the group $\Spin(d)$ which is the double cover of $\SO(d)$. (For instance, in $d \geq 3$ the homotopy groups are $\pi_1(\SO(d)) = \bZ_2$ and $\pi_1(\Spin(d))=0$.)

Before proceeding, it might be helpful to clarify some conceptual point.
Even if there are no fermions at all, it is possible to take the Lorentz symmetry group to be $\Spin(d)$ rather than $\SO(d)$. It is just a choice of data of the theory.
For example, when we consider a new physics model in a bottom up approach, we can freely choose a gauge group $G$ in the model building. The choice of $G$ is part of the data specifying the theory, and we can consider $G$ even if there are no matter fields (e.g., pure Yang-Mills).\footnote{If we take into account the completeness hypothesis~\cite{Polchinski:2003bq,Banks:2010zn,Harlow:2018tng}, the story here is more complicated. However, that hypothesis only requires very massive particles and hence may not contribute to low energy physics. } In the same way, in a quantum gravitational theory in a bottom up approach (rather than a top down approach such as string theory), 
the choice of the precise Lorentz group is part of the data of the theory under consideration. This choice affects e.g. gravitational path integrals as we briefly mention later. When there are fermions, the option $\SO(d)$ is simply not possible. 

Next let us consider the case that there is a CP symmetry. Without fermions, the Lorentz group (in Euclidean signature) including a CP symmetry can be taken to be $\O(d)$. 
It contains $\SO(d)$ as a subgroup. 
In the presence of fermions, the group $\O(d)$ is not possible and we need to take a double cover of $\O(d)$ which contains $\Spin(d)$ as a subgroup. 

It is known that there are two such groups, denoted by $\Pin^+(d)$ and $\Pin^-(d)$. The difference between them is as follows. First, recall that any element of $\O(d)$ can be written as
\beq
\tilde \sM \text{~~or~~} \tilde \sM  \tilde \sR_n,
\eeq
where $\tilde \sM \in \SO(d)$, and $\tilde \sR_n \in \O(d)$ which is explicitly given by the matrix $(\tilde \sR_n)^\mu_{\nu}=\delta^\mu_\nu -2n^\mu n_\nu$.
In $\Pin^\pm (d)$, there are two elements corresponding to each of $\tilde \sM$ and $\tilde \sR$. For $\tilde \sM$, let $\sM \in \Spin(d)$ be one element corresponding to $\tilde \sM$. The other element is given by $(-1)^\sF \sM$, where $(-1)^\sF \in \Spin(d)$ is ``the $2\pi$ rotation''. In the same way, for $\tilde \sR_n$, we have two corresponding elements. We denote one of them by $\sR_n$, and then the other is $(-1)^\sF \sR_n$. In $\O(d)$, we have $(\tilde \sR_n)^2=1$. Therefore, in a double cover of $\O(d)$, the square $(\sR_n)^2$ must be either $1$ or $(-1)^\sF$. The groups $\Pin^\pm(d)$ are characterized by the properties that they are double covers of $\O(d)$ such that
\beq
& \Pin^+(d) ~:~ (\sR_n)^2=1, \nonumber \\
& \Pin^-(d) ~: ~ (\sR_n)^2= (-1)^\sF.
\eeq
Both of them are relevant for string theory and condensed matter physics.

In the present paper, we focus on $\Pin^+(d)$ or its extension by a gauge group discussed later. The reason is that if we require $\sR_n$ to map a Weyl fermion to (the hermitian conjugate of) itself as in the case of a single Weyl fermion, we automatically have $\sR_n^2=1$ as discussed in the previous subsection (see \eqref{eq:pin+condition} and its general justification \eqref{eq:pin+condition2}). When we introduce a gauge group $G$, we will require that any Weyl fermion $\psi$ in an irreducible representation of the gauge group $G$ is mapped to (the hermitian conjugate of) itself under $\sR_n$. 
Under this requirement, we still have $\sR_n^2=1$ and hence the relevant group is $\Pin^+(d)$. 
On the other hand, to realize $\Pin^-(d)$ in four dimensions, we need to have an even number of fermions $(\psi_1, \psi_2, \cdots)$ as briefly mentioned at the end of Section~\ref{sec:singleneutral}. We do not consider such a case in the present paper.

\subsection{Inclusion of gauge group}

In Section~\ref{sec:singleneutral} we have discussed the CP transformation of a single neutral Weyl fermion. We now generalize it to the following case.
We introduce a gauge group which is denoted by $G$ (e.g., $G=\SU(3) \times \SU(2) \times \U(1)$), and we consider a Weyl fermion $\psi$ in an irreducible representation $\rho$ of $G$. We assume that $\psi$ is mapped to (the hermitian conjugate of) itself, rather than another fermion, under $\sR_n$. 

It is still possible to use the 4-component spinor notation by defining
\beq
\Psi = \begin{pmatrix} \psi_\alpha \\ \bar\psi^{\dot \alpha} \end{pmatrix}, \qquad  \gamma_5=\begin{pmatrix} 1 & 0 \\ 0 & -1 \end{pmatrix},
\eeq
where $\alpha=1,2$ and $\dot\alpha=1,2$ are appropriate 2-component spinor indices. 
The discussions of Section~\ref{sec:singleneutral} are still valid, and in particular we can use \eqref{eq:parity1} with $\alpha = \pm \i$ for the definition of the CP symmetry. 

However, the gauge transformation by $\sg \in G$ is now given by 
\beq
\Psi \to ( \rho(\sg) P_+ +  \rho^*(\sg) P_-) \Psi,
\eeq
where $P_\pm = \frac12 (1\pm \gamma_5)$ and $\rho^*$ is the complex conjugate representation of $\rho$. 
Because $P_\pm \gamma_5 \slashed{n} = \gamma_5 \slashed{n} P_\mp$, the transformation by $\sR_n$ does not commute with the gauge transformation of $G$.

This problem is of course solved by charge conjugation as follows. We assume that there is an automorphism 
\beq
\sigma : G \to G,\label{eq:automorphism}
\eeq
with the properties that $\sigma^2=1$ and
\beq
 \rho(\sigma(\sg)) =\rho^*(\sg), \label{eq:propertyofsigma}
\eeq
under an appropriate basis for the representation $\rho$. 
For instance, the automorphism $\sigma$ for the case $G=\SU(N)$ is simply given by complex conjugation $\sigma(\sg) = \sg^*$, where we regard $\sg \in \SU(N)$ as an $N \times N$ matrix.

The gauge field $A_\mu$ for the gauge group $G$ takes values in the Lie algebra of $G$, and there is a natural action of the automorphism $\sigma$ on the Lie algebra that is induced from the action on $G$. By using it, we can define the transformation of $\Psi$ and $A_\mu$ under $\sR_n$ by
\beq
\sR_n : \left\{ \begin{array}{l} A_\mu(x) \to A'_\mu(x) = (\delta^\nu_\mu -2n_\mu n^\nu) \sigma(A_\nu(x')) \\
\Psi(x)  \to \Psi'(x)= \alpha \gamma_5\slashed{n} \Psi(x') \label{eq:CPrep}
\end{array} \right.
\eeq
where $x'$ is given by \eqref{eq:xtransf}. Although we use the notation $\sR_n$, this transformation includes charge conjugation as well as parity. One may prefer to use the notation $\sC\sR_n$ rather than $\sR_n$, but it is just a matter of notation and we continue to use $\mathrm{R}_n$. 

The fermion Lagrangian
\beq
\cL_\Psi = -\frac12 \Psi^T C \slashed{D} \Psi
\eeq
is invariant under $\sR_n$, where the covariant derivative is given by
\beq
D_\mu \Psi = (\partial_\mu + \rho(A_\mu)P_+ + \rho^*(A_\mu) P_-) \Psi.
\eeq
Here we have taken the generators of the Lie algebra to be anti-hermitian, and absorbed the gauge coupling into the normalization of $A_\mu$. The gauge kinetic term  $\sim \frac{1}{4g^2}F_{\mu\nu} F^{\mu\nu}$ is also invariant under $\sR_n$. 

Now we can see that the precise symmetry group is given by
\beq
\Pin^+(d) \ltimes G \quad \text{with $d=4$}
\eeq
which is a semidirect product characterized by the following commutation relations between elements of $\Pin^+(d)$ and $G$ :
\beq
 \sM \sg &= \sg \sM  \quad ( \sM \in \Spin(d), ~ \sg \in G), \nonumber \\
 \sR_n \sg &=  \sigma(\sg) \sR_n. \label{eq:semidirect}
\eeq

There is some detail about representations of $\Pin^+(d)$, which would be important for odd dimensions (such as $d=3$) but is irrelevant for even dimensions (such as $d=4$). 
As we have seen in Section~\ref{sec:singleneutral}, the phase $\alpha$ in \eqref{eq:CPrep} is either $\i$ or $-\i$. These two signs give two representations of $\Pin^+(d)$. 
In these two representations, the action of $\sM \in \Spin(d)$ is the same, but the action of $\sR_n \in \Pin^+(d)$ differs by the sign. However, in $d=4$ (or in more general even dimensions), the two representations are equivalent since we can change $\Psi$ to a new field $\Psi'= \i \gamma_5 \Psi$ such that the sign of $\alpha$ is reversed. Therefore, for our purposes, we can just take $\alpha=\i$ without loss of generality.\footnote{
When we said that the sign of $\alpha$ is irrelevant, it is about a ``boundary theory'' in anomaly inflow. When there is a mass term, the change $\Psi'= \i \gamma_5 \Psi$ is accompanied by the change of the sign of the mass term. This has a significant effect for the ``bulk theory'' of the anomaly inflow~\cite{Witten:2015aba}.
} 
We remark that $(-1)^\sF \sR_n \in \Pin^+(d)$ acts on $\Psi$ by $-\alpha \gamma_5 \slashed{n}$ rather than $\alpha \gamma_5 \slashed{n} $, but this is due to the difference of group elements ($\sR_n$ or $(-1)^\sF \sR_n$) rather than the difference of representations.

We can also consider a scalar field $\phi$ in an irreducible representation $\rho$ of $G$.
For simplicity, we assume that $\rho$ is a complex representation and $\phi$ is a complex scalar.
Because of the charge conjugation, it must transform nontrivially under $\sR_n$ as
\beq
\sR_n : \phi(x) \to \phi^*(x'), \label{eq:scalartransf}
\eeq
where $\phi^*$ is the hermitian conjugate of $\phi$. 
A possible phase factor can be eliminated by a redefinition of $\phi$. (This is not necessarily the case if $\phi$ is real.)

\subsection{Summary of the group and representation}

Let us summarize the discussions so far. 
\begin{itemize}
\item If the CP symmetry is not considered, the symmetry group (in Euclidean signature) is assumed to be $\Spin(d) \times G$ (with $d=4$) where $\Spin(d)$ is the Lorentz symmetry and $G$ is the gauge group. 
\item We consider a Weyl fermion $\psi$ in an irreducible representation $\rho$ of $G$. 
\item The CP transformation is assumed to transform $\psi$ to (the hermitian conjugate of) itself. 
\item Under the above assumptions, the symmetry group is $\Pin^+(d) \ltimes G$ (with $d=4$). The semidirect product is characterized by \eqref{eq:semidirect} where $\sigma$ is the automorphism of $G$ such that \eqref{eq:propertyofsigma} is satisfied. Fields transform under the reflection along a spatial direction $n$ by \eqref{eq:CPrep} with $\alpha=\i$. 
\end{itemize}

\subsection{Gauging CP symmetry}\label{sec:gaugingCP}

For the purpose of studying global anomalies, it is necessary to understand what it means to gauge a CP symmetry. 
For this purpose, let us recall some standard facts. (Readers who are familiar with the standard mathematical description of gauge and gravitational fields by fiber bundles and spin structures of manifolds may skip the following few paragraphs. We do not explain all the details of these mathematical concepts; see e.g. \cite{Nakahara:2003nw}.)

Gauging an ordinary symmetry $G$ roughly means that we introduce a gauge field $A_\mu$ and impose gauge invariance. However, it is not the most precise description.
On a general spacetime, a gauge field configuration is mathematically described by a $G$-fiber bundle and its connection. The gauge field $A=A_\mu \d x^\mu$ is the connection of the fiber bundle. 
Consider two regions $U_\alpha$ and $U_\beta$ on the spacetime such that their overlap is nontrivial, 
$U_\alpha \cap U_\beta \neq \varnothing$. On these regions, we have connections $A_\alpha$ and $A_\beta$ represented by 1-forms (by choosing local trivializations of the fiber bundle). They are related by a transition function $\sg_{\alpha\beta} : U_\alpha \cap U_\beta \to G$ as $(A_\beta)_\mu = \sg_{\alpha\beta}^{-1} (A_\alpha )_\mu \sg_{\alpha\beta} + \sg_{\alpha\beta}^{-1} \partial_\mu \sg_{\alpha\beta}$. A general gauge field configuration is obtained by gluing various regions by using transition functions in this way. The consistency of the construction requires $\sg_{\alpha\beta}\sg_{\beta\gamma}=\sg_{\alpha\gamma}$ on $U_\alpha \cap U_\beta \cap U_\gamma$. 

Next we consider gauging the Lorentz symmetry. Roughly the gauge field for the Lorentz symmetry is the metric $g_{\mu\nu}$. However, even before introducing a metric, a general spacetime requires the concept of a manifold $X$. Consider two regions $U_\alpha$ and $U_\beta$ on $X$ such that their overlap is nontrivial, $U_\alpha \cap U_\beta \neq \varnothing$. On each of them, we have coordinate systems $x^\mu_\alpha$ and $ x^\mu_\beta$. On $U_\alpha \cap U_\beta$, one coordinate system is a collection of functions of the other, $x_\beta^\mu=x_\beta^\mu(x_\alpha)$. This is the analog of the transition function $\sg_{\alpha\beta}$ in the case of a gauge field. The metrics $(g_\alpha)_{\mu\nu}$ and $(g_\beta)_{\mu\nu}$ on $U_\alpha$ and $U_\beta$ are related by $(g_\beta)_{\mu\nu} = \displaystyle \frac{\partial x^\rho_\alpha}{\partial x^\mu_\beta} \frac{\partial x^\sigma_\alpha}{\partial x^\nu_\beta} (g_\alpha)_{\rho\sigma}$. This is the analog of $(A_\beta)_\mu = \sg_{\alpha\beta}^{-1} (A_\alpha )_\mu \sg_{\alpha\beta} + \sg_{\alpha\beta}^{-1} \partial_\mu \sg_{\alpha\beta}$. 

However, the above description of gauging the Lorentz symmetry is not the end of the story. When the Lorentz symmetry group is $\SO(d)$ (rather than $\Spin(d)$), the above description is almost sufficient. One additional point is that the manifold $X$ must be oriented, which in particular implies that the transition functions are such that
$\det( {\partial x_\alpha^\mu}/{\partial x_\beta^\nu} ) >0$. 
%\beq
%\det \left( \frac{\partial x_\alpha^\mu}{\partial x_\beta^\nu} \right) >0. 
%\eeq
This corresponds to the fact that the determinant of elements of $\SO(d)$ (as opposed to $\O(d)$) is $+1$ and hence positive.

A slight reformulation of the case of $\SO(d)$ is as follows. We introduce an orthonormal basis $e^\mu_a$ such that $e^\mu_a e^\nu_b g_{\mu\nu}=\delta_{ab}$. Instead of using $g_{\mu\nu}$, we can use $e^\mu_a$ since $g_{\mu\nu}=e^a_\mu e^b_\nu \delta_{ab}$ where $(e^a_\mu)$ is the inverse matrix of $(e^\mu_a)$. In this reformulation, we need to impose additional gauge invariance under the transformation $e^\mu_a \to e^\mu_a \tilde \sM^a_{~b}$. When the Lorentz symmetry group is $\SO(d)$, we take $\tilde \sM=( \tilde \sM^a_{~b})$ to have values in $\SO(d)$. Different regions $U_\alpha$ and $U_\beta$ are related by a transition function $\tilde \sM_{\alpha\beta}$ such that
\beq
(e_\beta)^\mu_a =  \frac{\partial x_\beta^\mu}{\partial x_\alpha^\nu} (e_\alpha)^\nu_b (\tilde \sM_{\alpha\beta})^b_{~a}. \label{eq:etransition}
\eeq
The transition functions $\tilde \sM_{\alpha\beta}$ among various regions form an $\SO(d)$ fiber bundle. 
The condition on the orientation discussed in the previous paragraph can be recovered as follows. On each region $U_\alpha$, we can choose a coordinate system $x_\alpha^\mu$ such that $\det ((e_\alpha)^\mu_a) >0$. Then, by taking the determinant of both sides of \eqref{eq:etransition}, we get $\det(\tilde \sM_{\alpha\beta})>0 \Longleftrightarrow \det( {\partial x_\beta^\mu}/{\partial x_\alpha^\nu} ) >0$. For $\SO(d)$ we have $\det(\tilde \sM_{\alpha\beta})=+1$ and hence $\det( {\partial x_\beta^\mu}/{\partial x_\alpha^\nu} ) >0$

When the Lorentz symmetry is $\Spin(d)$ rather than $\SO(d)$, a gravitational configuration is described not only by a manifold and a metric, but also a $\Spin(d)$ fiber bundle.
The $\Spin(d)$ fiber bundle is not an arbitrary one, but is constrained by the following conditions. First of all, its transition functions $\sM_{\alpha\beta}$ take values in $\Spin(d)$. There is the map $\pi : \Spin(d) \to \SO(d)$ representing the fact that $\Spin(d)$ is the double cover of $\SO(d)$. Then $\sM_{\alpha\beta}$ must be reduced to $\tilde \sM_{\alpha\beta}$, i.e., $\pi(\sM_{\alpha\beta}) =\tilde \sM_{\alpha\beta}$, where $\tilde\sM_{\alpha\beta}$ is the transition function for $e^\mu_a$ discussed in the previous paragraph. (The transition functions must of course satisfy the usual consistency condition $\sM_{\alpha\beta}\sM_{\beta\gamma}=\sM_{\alpha\gamma}$). For a given manifold $X$, such an $\Spin(d)$ bundle does not necessarily exist, and when it exists, it is not necessarily unique. A gravitational configuration is specified by $e^\mu_a$ and an $\Spin(d)$ bundle. The uplift of the $\SO(d)$ bundle to an $\Spin(d)$ bundle is called a spin structure, and a manifold with a spin structure is called a spin manifold.\footnote{A famous example is the case $X = S^1 \times Y_{d-1}$, where $Y_{d-1}$ is $(d-1)$-dimensional. Let $\tau $ be a coordinate of $S^1$ such that $\tau \sim \tau+\beta$, where $\beta$ is the circumference of $S^1$. We can impose either the antiperiodic  or periodic boundary condition on fermions (i.e., $\psi(\tau+\beta,y)=-\psi(\tau,y)$ or $\psi(\tau+\beta,y) = \psi(\tau,y)$). These are the example of different spin structures on $S^1$. If we regard $S^1$ as constructed from $[0,\beta]$ by gluing $\tau=\beta$ and $\tau=0$, the transition function between $\tau=\beta$ and $\tau=0$ is $(-1)^\sF$ or $1$, respectively.}
The connection on $\Spin(d)$ is just determined by the metric or $e^\mu_a$. Gravitational path integrals are the integrals over these configurations, including a sum over various spin manifolds and spin structures.

It should be now clear what it means to gauge a CP symmetry. First let us consider pure gravity without a gauge group $G$. For concreteness we discuss the case of $\Pin^+(d)$ although the case of $\Pin^-(d)$ is completely analogous. 
First, we consider the transitions like \eqref{eq:etransition} but we do not impose the condition $\det( \tilde \sM_{\alpha\beta})=1$. We allow more general transition functions valued in $\O(d)$. In particular, under the choice of a coordinate system such that $\det (e^\mu_a) >0$, we have the correspondence $\det( \tilde \sM_{\alpha\beta})=-1 \Longleftrightarrow \det( {\partial x_\alpha^\mu}/{\partial x_\beta^\nu} ) <0$ and hence the manifold may not be orientable.\footnote{A famous example of such a manifold in two dimensions is Klein bottle, described by $(x,y) \in \bR^2$ with the equivalence relations $(x+1,y) \sim (x,y)$ and $(x,y+1) \sim (-x,y)$.}. 
Next, we require that $\tilde \sM_{\alpha\beta}$ is uplifted to $\sM_{\alpha\beta} \in \Pin^+(d)$. Since $\Pin^+(d)$ is a double cover of $\O(d)$, we have the map $\pi : \Pin^+(d) \to \O(d)$, and the uplift $\sM_{\alpha\beta}$ is such that $\pi(\sM_{\alpha\beta}) =\tilde \sM_{\alpha\beta}$. 
We require $\sM_{\alpha\beta}$ to form a $\Pin^+(d)$ fiber bundle.

In the present paper, we are interested in the symmetry group $\Pin^+(d) \ltimes G$. 
This case is treated as follows. There is a group homomorphism $\pi : \Pin^+(d) \ltimes G \to \O(d)$ that is obtained by combining $\Pin^+(d) \ltimes G \to \Pin^+(d)$ (which just ``forgets'' $G$) and $\Pin^+(d) \to \O(d)$. A gauge and gravitational configuration is described by a $\Pin^+(d) \ltimes G $ fiber bundle with transition functions $\sL_{\alpha\beta}$
such that $\pi(\sL_{\alpha\beta})=\tilde \sM_{\alpha\beta}$, where $\tilde \sM_{\alpha\beta}$ are the same $\O(d)$ transition functions which appear in \eqref{eq:etransition}.
A connection of the bundle takes values in the Lie algebra $\so(d) \times \g$ of $\Pin^+(d) \ltimes G$. The $\g$ part of the connection is $A_\mu$, and the $\so(d)$ part is the one determined by the metric or $e^\mu_a$. 

For our later purpose, it is important to see how a scalar field in a representation $\rho$ of $G$ behaves in the background of a general $\Pin^+(d) \ltimes G$ bundle. 
Let $\phi_\alpha$ and $\phi_\beta$ be scalar fields in the regions $U_\alpha$ and $U_\beta$.
If the transition function between $U_\alpha$ and $U_\beta$ is given by $\sg_{\alpha\beta} \in G \subset \Pin^+(d) \ltimes G$, we have
\beq
\phi_\beta = \rho(\sg_{\beta\alpha})\phi_\alpha. \label{eq:str1}
\eeq
On the other hand, if the transition function is given by $\sR_{\alpha\beta} \in \Pin^+(d) \subset \Pin^+(d) \ltimes G$ such that $\det(\tilde \sR_{\alpha\beta})=-1$, we have
\beq
\phi_\beta=\phi_\alpha^*. \label{eq:str2}
\eeq
A transition function $\sM_{\alpha\beta}$ whose value is in $\Spin(d)$ acts trivially. A general transition function is a combination of these. 
Thus, the scalar field takes values in a vector bundle $E$ that is associated to the $\Pin^+(d) \ltimes G$ bundle, with the following properties. To be concrete, we assume that the representation $\rho$ is complex. Each fiber $E_x$ (where $x \in X$ is a point so that $\phi(x) \in E_x$) is a vector space whose complex dimension is $\dim_{\bC} \rho$ and real dimension is $\dim_{\bR} \rho=2\dim_{\bC} \rho$. For instance, if $G=\SU(N)$ and $\rho$ is the defining, $N$-dimensional representation of $\SU(N)$, then $\dim_{\bC} \rho=N$ and $\dim_{\bR} \rho=2N$. Globally, the vector bundle $E$ is constructed by gluing various regions by transition functions like \eqref{eq:str1} and \eqref{eq:str2}.

\section{Anomalies }\label{sec:anomalies}

\subsection{Anomaly inflow}

Here we briefly review the formulation of anomalies in terms of anomaly inflow. For simplicity, we consider the path integral without any operator insertions, but the following discussions do not essentially change even if we include operators. The explanations in this subsection are only heuristic; see \cite{Witten:2019bou} and references therein for more systematic, nonperturbative treatments. 

Let $\cA_X =(X,e^\mu_a, A_\mu)$ represent all the information of the spacetime manifold $X$ (where $\dim X=d$), the metric $e^\mu_a$ and the gauge field $A_\mu$. Let $\psi$ be the chiral fermion which we are interested in. 
The Euclidean path integral is naively given by
\beq
\cZ &= \int [D\cA_X D\psi] e^{-S} = \int [D \cA_X] \cZ_\psi(\cA_X), \nonumber \\
\cZ_\psi(\cA_X) &=  \int [ D\psi] e^{-S},
\eeq
where $S$ is the Euclidean action. Then, perturbative anomalies may be described as follows. Under an infinitesimal gauge transformation $\cG$ which is schematically written by 
$\cA_X  \to \cA_X^\cG$, the fermion partition function $\cZ_\psi(\cA_X)$ changes as
\beq
\cZ_\psi(\cA_X^\cG) = \cZ_\psi(\cA_X) \exp\left( - 2\pi \i \int_{X} I^{(1)}_{d}(\cA_X, \cG) \right) 
\eeq
where $I^{(1)}_{d}(\cA_X, \cG) $ is the perturbative anomaly of $\psi$. It is a differential $d$-form (the details are not so important). By the anomaly descent equation (see e.g., \cite{Weinberg:1996kr} for textbook accounts), there is a Chern-Simons $(d+1)$-form $I_{d+1}^{(0)}(\cA_X)$ such that 
\beq
I_{d+1}^{(0)}(\cA_X^\cG) -I_{d+1}^{(0)}(\cA_X) = \d I^{(1)}_{d}(\cA_X, \cG)
\eeq
where $\d$ on the right-hand side means total derivative (or exterior derivative in differential forms). 

\begin{figure}
\centering
\includegraphics[width=0.3\textwidth]{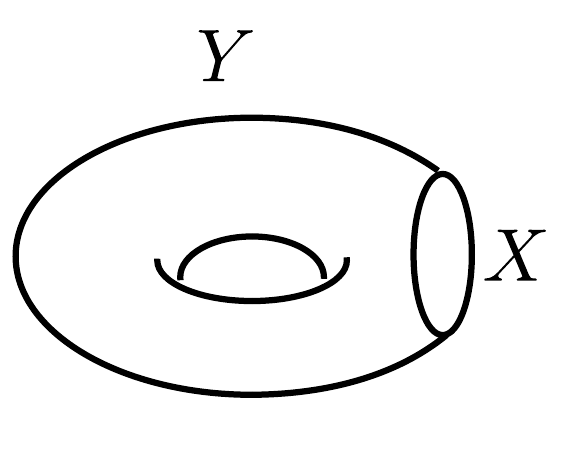}
\caption{An example of a $(d+1)$-dimensional manifold $Y$ and its $d$-dimensional boundary $X$. In this example, $d=1$, $X=S^1$, and $Y$ is a torus with a disk removed.   \label{fig:1}}
\end{figure}

Perturbative anomaly inflow~\cite{Callan:1984sa} is described as follows.
We take an auxiliary $(d+1)$-dimensional manifold $Y$ which has boundary $X$,
\beq
\partial Y = X
\eeq
where $\partial$ on the left-hand side means taking the boundary. 
See Figure~\ref{fig:1} for an example of $X$ and $Y$.
We also extend $e^\mu_a$ and $A_\mu$ to $Y$ in an arbitrary way, and denote them by $ \cA_Y=(Y,e^\mu_a,A_\mu)$. Then we modify the fermion partition function as
\beq
 \tilde \cZ_\psi ( \cA_Y) = \cZ_\psi(\cA) \exp\left( 2\pi \i \int_{Y} I^{(0)}_{d+1}(\cA_Y) \right).
\eeq
By the Stokes theorem $\int_{Y} \d I^{(1)}_{d}(\cA_Y, \cG) = \int_{X}I^{(1)}_{d}(\cA_X, \cG) $, the modified partition function $ \tilde \cZ_\psi ( \cA_Y)$ can be seen to be invariant under the gauge transformation $\cA_Y  \to \cA_Y^\cG$,
\beq
 \tilde \cZ_\psi ( \cA_Y^\cG) = \tilde \cZ_\psi ( \cA_Y).
\eeq

\begin{figure}
\centering
\includegraphics[width=0.9\textwidth]{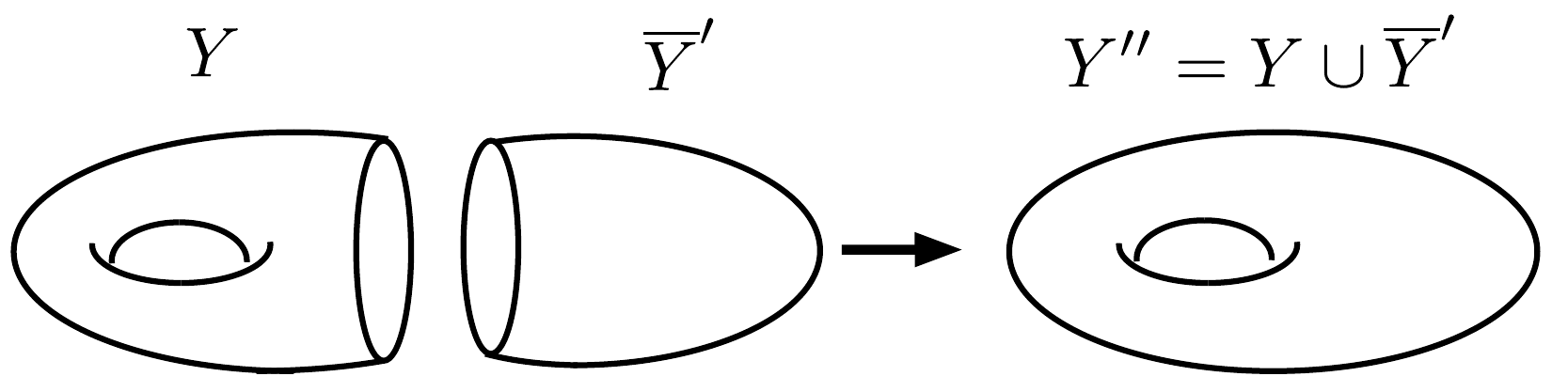}
\caption{Gluing of $Y$ and (the orientation reversal $\overline{Y}'$ of) $Y'$ along the common boundary $\partial Y = \partial Y' =X$ to obtain the closed manifold $Y''=Y \cup \overline{Y}'$.   \label{fig:2}}
\end{figure}

Anomalies in general are formulated by using anomaly inflow. We make the fermion partition function gauge invariant by anomaly inflow mechanism. However, the cost for it is that we need to introduce an auxiliary $(d+1)$-dimensional manifold $Y$ which is not physical in the current particle physics context.\footnote{In some setup of condensed matter physics, string theory, and topological solitons in particle physics, the higher dimensional manifold can have physical meanings depending on the context. In the present paper, we are mainly interested in particle phenomenology in $d=4$ dimensions, and in this context the 5-dimensional manifold $Y$ does not have a physical meaning.} Suppose we take another manifold $Y'$ whose boundary is the same $X$, and extend the fields to it. Let $\cA_{Y'}=(Y',e'^\mu_a, A'_\mu) $ be the total information on $Y'$. Then, the dependence of $\tilde \cZ_\psi$ on the auxiliary data can be seen by taking the ratio
\beq
\frac{\tilde \cZ_\psi ( \cA_Y)}{\tilde \cZ_\psi ( \cA_{Y'})} := \cZ_{\rm anomaly}(\cA_{Y''}),\label{eq:theanomaly}
\eeq
where $Y''$ is the closed manifold that is obtained by gluing $Y$ and (the orientation reversal $\overline{Y}'$ of) $Y'$ along the common boundary $X$; see Figure~\ref{fig:2} for the situation. The right-hand side $\cZ_{\rm anomaly}(\cA_{Y''})$ is, at the perturbative level, given by 
\beq
\cZ_{\rm anomaly}(\cA_{Y''}) \xrightarrow{\rm perturbative} \exp\left( 2\pi \i \int_{Y''} I^{(0)}_{d+1}(\cA_{Y''}) \right) . \label{eq:panomaly}
\eeq

The quantity $\cZ_{\rm anomaly}(\cA_{Y''})$ on the right-hand side of \eqref{eq:theanomaly} is the proper definition of anomaly, by the following argument. Suppose that $\cZ_{\rm anomaly}(\cA_{Y''})$ is always unity, $\cZ_{\rm anomaly}(\cA_{Y''})=1$, for arbitrary $\cA_{Y''}$. Then, \eqref{eq:theanomaly} implies that for any $Y$ and $Y'$ we have $\tilde \cZ_\psi ( \cA_Y) = \tilde \cZ_\psi ( \cA_{Y'})$. This means that $\tilde \cZ_\psi$ actually does not depend on the choice of the auxiliary data $\cA_Y$ which is unphysical. It only depends on the original data $\cA_X$ on the $d$-dimensional manifold $X$, and hence we can regard it as a function of $\cA_X$,
\beq
\tilde \cZ_\psi = \tilde \cZ_\psi ( \cA_X) \quad \text{if} ~ \cZ_{\rm anomaly}(\cA_{Y''})=1 ~(\forall Y'').
\eeq
This $\tilde \cZ_\psi $ can be taken as the definition of the fermion partition function. By construction, it is gauge invariant.

\subsection{Perturbative anomaly}\label{sec:perturbative}
Let us consider perturbative anomalies in a little more detail. The presence of a CP symmetry does not change perturbative anomalies, but there is a consistency check which we discuss below.

A perturbative anomaly is described by an anomaly polynomial $I_{d+2}$ which is related to $I^{(0)}_{d+1}$ by\,\footnote{\eqref{eq:polyCS} is valid only for topologically trivial gauge field configurations. For topologically nontrivial configurations, $I_{d+2}$ is well-defined but not necessarily an exact form, and $I^{(0)}_{d+1}$ is not necessarily well-defined. Instead of $I^{(0)}_{d+1}$, we need to use the Atiyah-Patodi-Singer (APS) $\eta$-invariant.}
\beq
I_{d+2} = \d I^{(0)}_{d+1}. \label{eq:polyCS}
\eeq
$I_{d+2}$ is more explicitly given as follows.
Let $F=\frac12 F_{\mu\nu} \d x^\mu \wedge \d x^\nu$ be the field strength regarded as a 2-form, which is given in terms of the gauge field $A=A_\mu \d x^\mu$ by $F=\d A+A \wedge A$ (or more explicitly $F_{\mu\nu}=\partial_\mu A_\nu - \partial_\nu A_\mu +[A_\mu, A_\nu]$). Let $R=(\frac12 R^{\mu}_{~\nu\rho\sigma} \d x^\rho \wedge \d x^\sigma)$ be the Riemann curvature regarded as a matrix-valued 2-form. The anomaly polynomial $I^\rho_{d+2} $ of a positive chirality Weyl fermion $\psi$ in a representation $\rho$ is given by
\beq
I^\rho_{d+2} = \left [ \tr_\rho \exp\left( \frac{\i F}{2\pi} \right) \hat{A}(R) \right]_{d+2} \label{eq:rhopoly}
\eeq
where $\hat{A}(R)$ is the $\hat{A}$-genus (whose details are not so important in the current paper), the trace is taken in the representation $\rho$, and the subscript $d+2$ on the right-hand side means that we take the $(d+2)$-form part. 

More explicitly, for $d=4$ we have
\beq
I^\rho_{6}  = \frac{1}{3!} \tr_\rho \left( \frac{\i F}{2\pi} \right)^3 -\frac{1}{24}p_1(R)\tr_\rho \left( \frac{\i F}{2\pi} \right). \label{eq:I6}
\eeq
where $p_1(R) = -\frac{1}{2} \tr (\frac{R}{2\pi})^2$ is the first Pontryagin class. The first term represents the triangle anomaly of the type $G$-$G$-$G$, and the second term represents $G$-$\text{(gravity)}$-$\text{(gravity)}$.

Geometrically, the anomaly polynomial $I_{d+2}$ can be used as follows. We mentioned that the anomaly is given by \eqref{eq:panomaly} at the perturbative level. Now, suppose that there exists a $(d+2)$-dimensional manifold $Z$ whose boundary is $Y''$ (i.e., $\partial Z=Y''$), and on which the fields are extended. We denote the information of the fields on $Z$ as $\cA_Z$. Then, by Stokes theorem, we get
\beq
\int_{Y''} I^{(0)}_{d+1}(\cA_{Y''})= \int_{Z} I_{d+2}(\cA_{Z}). \label{eq:d+2int}
\eeq
The right-hand side has the advantage that $I_{d+2}$ is manifestly gauge invariant and hence its integral can be defined even for topologically nontrivial field configurations. On the other hand, the expression on the left-hand side is valid only for perturbative purposes when the fields are topologically trivial. The correct nonperturbative statement which is valid for nontrivial topologies is that if $Y''=\partial Z$, then we have
\beq
\cZ_{\rm anomaly}(\cA_{Y''}) = \exp\left( 2 \pi \i \int_{Z} I_{d+2}(\cA_{Z}) \right) \quad \text{if}~ Y'' = \partial Z.  \label{eq:d+2int2}
\eeq

The anomaly polynomial is the same regardless of whether there is a CP symmetry or not. However, there is an important consistency check as follows.\footnote{
A more general mathematical structure is discussed in \cite{Yamashita:2021cao}.
}
Recall that we need transition functions like \eqref{eq:etransition} between two different regions $U_\alpha$ and $U_\beta$ on a manifold. As in Section~\ref{sec:gaugingCP}, we use the convention that $\det(e_\alpha)>0$ for any $\alpha$.
In $(d+2)$ dimensions, the relevant Lorentz group is $\Pin^+(d+2)$.
When the transition function $\sM_{\alpha\beta}$ of $\Pin^+(d+2)$ is such that its reduction 
$\tilde \sM_{\alpha\beta}$ to $\O(d+2)$ has 
$\det(\tilde \sM_{\alpha\beta})<0$, then the orientation is changed under the coordinate change, $\det(\partial x^\mu_\beta/\partial x^\nu_\alpha)<0$. 
In the definition of the integral \eqref{eq:d+2int}, we need orientation (because the integral of a differential form involves the totally antisymmetric tensor $\epsilon^{\mu_1 \cdots \mu_D}$ in $D$ dimensions). Therefore, one might think that the integral \eqref{eq:d+2int} is not well-defined because of the sign change under the transitions between different regions.

Fortunately, this sign change is canceled by another sign change.
The gauge field transforms under the automorphism $\sigma$ which exchanges $\rho$ and its complex conjugate representation $\rho^*$. We have
\beq
 \tr_{\rho^*} \left( \frac{\i F}{2\pi} \right)^n = (-1)^n   \tr_\rho \left( \frac{\i F}{2\pi} \right)^n .
\eeq
Therefore, for odd $n$, the sign coming from $\sigma$ and the orientation reversal cancel with each other. 
This is precisely what happens in $d=4$ (or more generally $d \in 4\bZ$), as can be seen from the expression \eqref{eq:I6} (or more generally \eqref{eq:rhopoly}) for $I^\rho_{d+2}$. By this mechanism, the integral \eqref{eq:d+2int} is well-defined even if the manifold $Z$ is not orientable.

\subsection{Global anomalies}
Global anomalies are nonperturbative anomalies in the absence of perturbative ones. In the rest of the paper, we assume that perturbative anomalies are absent, which means that the anomaly polynomial vanishes,
\beq
I_{d+2}=0.
\eeq
In such cases, anomalies are characterized as follows.

Recall the basic formula \eqref{eq:theanomaly}.
We have reviewed this formula in the case of perturbative anomalies, but it is known to be valid even nonperturbatively~\cite{Witten:1999eg,Witten:2015aba,Witten:2019bou}. 
However, the right-hand side of that formula, $\cZ_{\rm anomaly}(\cA_{Y''})$, cannot be written as a simple integral. 

For notational simplicity, we omit the double prime on $Y$ and simply write the anomaly as $\cZ_{\rm anomaly}(\cA_{Y})$ with $\partial Y = \varnothing$ (i.e., the empty boundary). A compact manifold without boundary is called a closed manifold. Thus, $Y,Y',\cdots$, which appear below are closed manifolds. 

\begin{figure}
\centering
\includegraphics[width=0.5\textwidth]{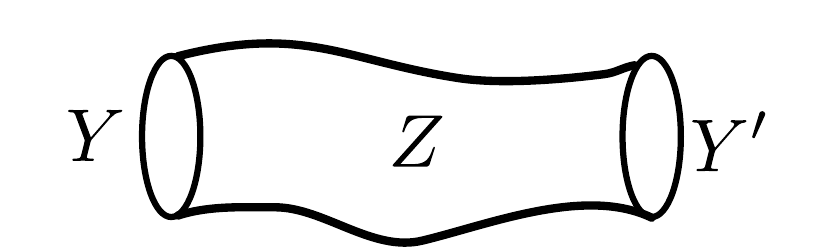}
\caption{A $(d+2)$-dimensional manifold $Z=Z$ whose boundary is $Y$ and $Y'$.   \label{fig:3}}
\end{figure}

One can show that $\cZ_{\rm anomaly}(\cA_{Y})$ is a topological invariant (or more strongly, bordism invariant) in the absence of perturbative anomalies. The argument is as follows. Suppose that two field configurations $\cA_Y$ and $\cA_{Y'}$ can be continuously deformed to each other. Then we can consider a $(d+2)$-dimensional manifold $Z$ as in Figure~\ref{fig:3} that realizes the  continuous deformation. The value of $\cZ_{\rm anomaly}$ on $\cA_Y$ and $\cA_{Y'}$ are related by
\beq
\frac{\cZ_{\rm anomaly}(\cA_{Y'})}{\cZ_{\rm anomaly}(\cA_{Y})} =\exp \left( 2\pi \i \int_{Z} I_{d+2} (\cA_{Z}) \right) .\label{eq:bordisminv}
\eeq
This equation is a version of \eqref{eq:d+2int2}.
When the anomaly polynomial $I_{d+2} (\cA_{Z})$ vanishes, we obtain $\cZ_{\rm anomaly}(\cA_{Y'})=\cZ_{\rm anomaly}(\cA_{Y})$. This is the desired topological invariance. 

In general studies of global anomalies, the property called bordism invariance is very important~\cite{Kapustin:2014tfa,Kapustin:2014dxa,Freed:2016rqq,Yonekura:2018ufj}. Suppose that there exists a $(d+2)$-manifold $Z$ and extension of the fields to it such that $\partial Z = Y$. When $I_{d+2} (\cA_{Z})=0$, \eqref{eq:d+2int2} gives
\beq
\cZ_{\rm anomaly}(\cA_{Y})=1 \quad \text{if~~} \exists Z\text{~such that~}  Y=\partial Z.
\eeq
This property is the bordism invariance. 
The bordism invariance of global anomalies is very powerful in general, and we will use it later in our discussions.

Throughout the present paper, we only discuss the case that global anomalies vanish. For several examples in other than four dimensions in which CP symmetry really gives nontrivial global anomalies $\cZ_{\rm anomaly}(\cA_{Y}) \neq 1$, see \cite{Witten:2015aba}.

\subsection{The absence of anomalies of the CP symmetry}
The main result in this paper is the following:
\begin{itemize}
    \item Consider the case $d=4$. Suppose that the gauge group $G$ is connected and simply connected, $\pi_0(G)=\pi_1(G)=0$. Suppose moreover that there is no anomaly before gauging the CP symmetry. Then, there is no new anomaly for the symmetry $\Pin^+(d) \ltimes G$ after gauging the CP symmetry.
\end{itemize}
As a corollary, the standard model matter content has no anomalies under gauging the CP. This is because the $\SU(5)$ grand unified theory (GUT) has no global anomalies before gauging the CP~\cite{Freed:2006mx,Garcia-Etxebarria:2018ajm,Witten:2019bou}, and the standard model gauge group can be embedded in $\SU(5)$ which satisfies $\pi_0(\SU(5))=\pi_1(\SU(5))=0$. Another consequence is that what is usually called the ``$\SO(10)$ GUT'', which should be more precisely called as $\Spin(10)$ GUT, is also anomaly free under gauging the CP. We will discuss string theory realization of it in Section~\ref{sec:string}.

Our purpose in this and the next subsection is to show the above result.\footnote{We do not try to make the argument mathematically rigorous, but we believe that it is possible to do so in a straightforward way, in the sense of computing the relevant bordism groups $\Omega^{\Pin^+ \ltimes G}_5({\rm pt})$.} The argument is completely analogous to that given in Section~4.4 of \cite{Witten:2019bou}, with only minor modifications.
In this subsection we discuss the case of $G=\SU(N)$ and in particular the $\SU(5)$ GUT, and in the next subsection we treat more general cases.

Consider the case $G=\SU(N)$. Our strategy is to use the topological invariance of $\cZ_{\rm anomaly}(\cA_{Y})$ discussed in the previous subsection to reduce to the case of $G=\SU(2)$.

What we want to show is that $\cZ_{\rm anomaly}(\cA_{Y})=1$ for arbitrary $\cA_Y=(Y,e^\mu_a,A_\mu)$ (where $\partial Y = \varnothing$). In the absence of perturbative anomalies, $\cZ_{\rm anomaly}$ is a topological invariant and hence explicit metrics and connections are irrelevant. Only the topology of the manifold $Y$ and the fiber bundle on it are important. 

We consider the vector bundle $E$ whose transition functions are given by \eqref{eq:str1}, \eqref{eq:str2} and their combinations. In particular, for the purpose of the following discussions, we take the representation $\rho$ to be the defining ($N$-dimensional) representation of $\SU(N)$. 

We take a generic-enough section $\phi$ of the vector bundle $E$ on $Y$. Physically, it might be helpful to imagine that $\phi$ is a scalar field configuration on $Y$. To aid the intuition, we sometimes call $\phi$ a scalar field. However, we emphasize that $\phi$ is introduced just for the purpose of technical (mathematical) discussions. 

If the configuration $\phi$ is generic enough, it is nonzero everywhere for $N>2$, by the following reason. The points $x \in Y$ on which $\phi$ vanishes is characterized by the equation $\phi(x)=0$. Notice that $\phi$ has $N$ complex components, or $2N$ real components. Therefore the equation $\phi(x)=0$ actually contains $2N$ real equations for each component. On the other hand, the number of variables $x$ is $d+1$, which is the dimension of $Y$. If $2N>d+1$, the number of equations is greater than the number of variables. Therefore, there is no solution if $\phi$ is generic enough, and in that case $\phi$ is everywhere nonzero. 

Let us take such an everywhere nonzero $\phi$. By replacing $\phi \to \phi/|\phi|$ if necessary, we may assume that it has unit length on each point of $Y$. Then, we can reduce the vector bundle $E$ as
\beq
E = E' \oplus \underline{\mathbb C} \label{eq:split}
\eeq
where $\underline{\mathbb C}$ means the trivial bundle spanned by the direction of $\phi$, and $E'$ is the  subbundle that is orthogonal to $\phi$. In particular, $E'$ has rank $N-1$. 

By $\SU(N)$ gauge transformations on each region $U_\alpha$, we can take a basis on $U_\alpha$ such that 
\beq
\phi = \begin{pmatrix} 0 \\ \vdots \\ 0 \\ 1 \end{pmatrix} \quad \text{(after $\SU(N)$ gauge transformations)}.
\eeq
Then the splitting \eqref{eq:split} is manifest.
The transition functions in the above basis of the fiber bundle are inside $\SU(N-1)$ (or more precisely $\Pin^+(d+1) \ltimes \SU(N-1)$). Thus the bundle is reduced from $G=\SU(N)$ to $G=\SU(N-1)$ by the above procedure. Intuitively, one might think that ``the $\SU(N)$ gauge group is Higgsed to $\SU(N-1)$ by the scalar $\phi$ without any topological obstruction''. (The case with topological obstruction will be discussed in the next subsection for $\SU(2)$.)

By repeating the above procedure, the bundle is reduced to the case of $\SU(2)$. Therefore the problem is reduced to the case of $G=\SU(2)$. The only thing we need to consider is the anomaly under the subgroup $\SU(2) \subset \SU(N)$, or more precisely $\Pin^+(d) \ltimes \SU(2)$. 

Now we can immediately show that the $\SU(5)$ with chiral fermions in the representation ${\bf 5} \oplus \overline{\bf 10}$ (i.e., one family of the $\SU(5)$ GUT) is anomaly free, as follows. Under the subgroup $\SU(2) \subset \SU(5)$, the representation is reduced as
\beq
{\bf 5} \oplus \overline{\bf 10} \xrightarrow{  \SU(2) \subset \SU(5)} (4 \times {\bf 2} )\oplus (7\times {\bf 1}).
\eeq
For an even number of fermions in the doublet representation ${\bf 2}$, we can add $\SU(2)$-invariant mass terms. These mass terms also preserve the CP symmetry (or more precisely the $\Pin^+(d)$ symmetry) as can be shown as in Section~\ref{sec:CP}. Any massive fermions do not contribute to anomalies. Also, we can add mass terms to any fermions in the singlet ${\bf 1}$ in a CP-invariant way. We conclude that the original ${\bf 5} \oplus \overline{\bf 10}$ does not have global anomalies under $\Pin^+(d) \ltimes \SU(5)$. 

\subsection{More details}
Here we discuss more general cases. This subsection assumes familiarity with some relevant mathematical facts.

\paragraph{The case of $\SU(N)$.} In the previous subsection, we have reduced the problem from $\SU(N)$ to $\SU(2)$ and given an elementary argument for ${\bf 5} \oplus \overline{\bf 10}$. The case of a more general matter content is argued as follows. 

For $\SU(2)$, we again consider a generic section (or a configuration of a scalar field) $\phi$.
In this case, $\phi$ has $2N=4$ real components, and the dimension of the manifold $Y$ is $d+1=5$. 
Then, even if $\phi$ is generic enough, we expect to have points $x \in Y$ where $\phi(x)=0$.
If $\phi$ is generic enough, the set 
\beq
M = \{ x \in Y\,|\, \phi(x)=0\},
\eeq
is a smooth submanifold of $Y$ with dimension $\dim M = d+1 - 2N$.\footnote{This is a consequence of the transversality theorem, or ultimately Sard's theorem.} In particular, $M$ is empty if $d+1<2N$; this fact was used in the previous subsection.

The submanifold $M$ may be regarded as ``topological obstruction to Higgsing $\SU(N)$ to $\SU(N-1)$''. In the case of our interest ($d+1=5$), there is no obstruction for $N>2$ as in the previous subsection. For $N=2$, the submanifold $M$ has dimension $\dim M=1$. We need to take care of it.

$M$ is a closed manifold. Any closed manifold of dimension 1 is (a disjoint union of) circles. Without loss of generality, we can simply assume $M=S^1$ for the purposes of the following discussions, because we can treat each circle separately.

An important point is that there is no orientation reversal in the neighborhood of $M$. The reason is as follows. 
A tubular neighborhood of $M=S^1$ in $Y$ is a disk bundle over $S^1$ with fiber the disk $D^4$. Then the only possibility for orientation reversal is that the orientation of $D^4$ is flipped as we go around the $S^1$. However, such a flip does not occur. It is possible to take an explicit coordinate system for $D^4$ by using $\phi$. For $N=2$, we can write
\beq
\phi = \begin{pmatrix} \phi_1 + \i \phi_2 \\ \phi_3 + \i \phi_4 \end{pmatrix},
\eeq
where $\phi_1 ,\ldots, \phi_4$ are real. The fact that $\phi$ is generic enough (or more precisely the transversality of $\phi$ to the zero section) implies that $\phi_1 ,\ldots, \phi_4$ form a good coordinate system for $D^4$ in the neighborhood of $M$. We can define the orientation of $D^4$ by 
$\d \phi_1 \wedge \d \phi_2 \wedge \d \phi_3 \wedge \d \phi_4 $. This orientation is invariant under the transformations \eqref{eq:str1} and \eqref{eq:str2}. Therefore, there is no orientation flip of the fiber $D^4$. 

Without orientation flip, a disk bundle over $S^1$ is topologically trivial and hence the tubular neighborhood of $M$ is of the form $  D^4 \times S^1$. We can interpret the situation by saying that ``there is an $\SU(2)$ instanton on $D^4$'', as can be seen as follows. Where $\phi \neq 0$, we can define an $\SU(2)$-valued function by
\beq
\sg = \frac{1}{|\phi|} \begin{pmatrix} \phi_1 + \i \phi_2 & -\phi_3 + \i \phi_4 \\ \phi_3 + \i \phi_4 & \phi_1 - \i \phi_2\end{pmatrix} \in \SU(2) .
\eeq
The gauge transformation of $\phi$ by $\sg$ gives
\beq
\sg^{-1} \phi = \begin{pmatrix}  |\phi| \\ 0 \end{pmatrix} 
\eeq
which is nonzero on $Y \setminus (D^4 \times S^1)$ (i.e., the complement of $D^4 \times S^1$ on $Y$). Therefore, the $\SU(2)$ bundle is reduced to $\SU(1)$ outside of $D^4 \times S^1$, which is trivial. If we use $\sg$ as a transition function on $\partial D^4 =S^3$, it gives an element $1 \in \bZ \simeq \pi_3(\SU(2))$. Thus, an ''instanton'' is localized on $D^4$, which then goes around $S^1$. 

Now we use a consequence of bordism invariance of global anomalies. We first prepare a manifold $  S^4 \times S^1$ on which $\SU(2)$ bundle is trivial. We cut it into two $ D^4 \times S^1$'s, where there is no instanton. 

Next we take the tubular neighborhood $D^4_{\rm instanton} \times S^1$ of $M$ on $Y$ discussed above, where the subscript ``${\rm instanton}$'' is put to indicate that the $D^4_{\rm instanton}$ contains an instanton. 
By removing $D^4_{\rm instanton} \times S^1$ from $Y$, we cut $Y$ into $D^4_{\rm instanton} \times S^1$ and
$Y \setminus D^4_{\rm instanton} \times S^1$. 

Bordism invariance implies that we can freely cut and glue manifolds (see e.g. \cite{Yonekura:2018ufj} for details). We have
\beq
&Y \sqcup ( S^4 \times S^1) \sim \tilde Y \sqcup ( S^4_{\rm instanton} \times S^1),
\eeq 
where $\sqcup$ means disjoint union, $\sim$ means that two sides are bordant, $\tilde Y$ is a new manifold in which the instanton is removed and in particular we can take an everywhere nonzero section $\phi$ on $\tilde Y$, and $S^4_{\rm instanton}$ is an $S^4$ with an instanton. Therefore, the bordism invariance of global anomalies implies that
\beq            
\cZ_{\rm anomaly}(\cA_{Y}) \cZ_{\rm anomaly}(\cA_{S^4 \times S^1})=
\cZ_{\rm anomaly}(\cA_{\tilde Y}) \cZ_{\rm anomaly}(\cA_{S^4_{\rm instanton} \times S^1}).
\eeq

We can finally show the absence of new anomalies under the CP. The $\tilde Y$ has a everywhere nonzero section $\phi$ and hence the gauge group is reduced to $\SU(1)$, which is trivial. Therefore, as far as $\tilde Y$ is concerned, we can neglect the gauge group $G$ and only consider pure $\Pin^+$ manifolds. A neutral Weyl fermion can have a mass term which is invariant under the CP (see Section~\ref{sec:CP}), and hence there should be no anomaly. This is consistent with the fact that $\Omega^{\Pin^+}_5({\rm pt})=0$. (See e.g. tables in \cite{Kapustin:2014dxa,Garcia-Etxebarria:2018ajm}.)
We conclude that
\beq
\cZ_{\rm anomaly}(\cA_{\tilde Y})=1.
\eeq
On the other hand, the manifold $ S^4_{\rm instanton} \times S^1$ is orientable, and hence the CP is irrelevant for this part. Our initial assumption that there is no anomaly before gauging the CP implies that
\beq
\cZ_{\rm anomaly}(\cA_{ S^4_{\rm instanton} \times S^1})=1.
\eeq
We also have $\cZ_{\rm anomaly}(\cA_{ S^4 \times S^1} )=1$. Therefore, we obtain
\beq
\cZ_{\rm anomaly}(\cA_{ Y})=1.
\eeq
This is the desired result that there is no new anomaly under gauging the CP.

\paragraph{The case of more general $G$.} 

We only give a brief discussion of the case for a general $G$ with $\pi_0(G)=\pi_1(G)=0$. Familiarity with the obstruction theory reviewed in \cite{Witten:1985bt} is assumed. 

Without loss of generality, we can assume that $G$ is simple. If it is semisimple, we can just consider each simple factor in the following discussions. 

The question is whether we can reduce $G$ to $\SU(2)$. In the case of $\mathrm{SU}(N)$, we have reduced $\mathrm{SU}(N)$ to $\mathrm{SU}(N-1)$ by using a section $\phi$ with $|\phi|=1$. For a general $G$ and its subgroup $H$, the analogous reduction from $G$ to $H$ is done as follows. For a given $\mathrm{Pin}^+(d) \ltimes G$ bundle, we can consider the associated bundle whose fiber is $\mathrm{Pin}^+(d) \ltimes G/\mathrm{Pin}^+(d) \ltimes H= G/H$. If there exists a global section of this bundle, then the original $\mathrm{Pin}^+(d) \ltimes G$ bundle can be reduced to a $\mathrm{Pin}^+(d) \ltimes H$ bundle. 

To show the existence of a section, we take a triangulation of $Y$. Let $Y^{(k)}$ be the $k$-skeleton that consists of simplices of dimensions less than or equal to $k$. 
Suppose that we have already found a section on $Y^{(k)}$, and we ask whether it can be extended to $Y^{(k+1)}$.

The $(k+1)$-skeleton $Y^{(k+1)}$ is obtained from $Y^{(k)}$ by attaching $(k+1)$-simplices to it. Let $\sigma$ be a $(k+1)$ simplex. A simplex is topologically trivial (contractible) and hence the bundle can be trivialized on $\sigma$. We already have a section on $\partial \sigma \subset Y^{(k)}$, which is just a function $\partial \sigma \to G/H$ since the bundle is trivialized on $\sigma$. We want to extend it to a function $\sigma \to G/H$. This is guaranteed to be possible if $\pi_k(G/H)=0$, because $\partial \sigma \simeq S^k$ and $\sigma \simeq D^{k+1}$. 

There is a long exact sequence
\begin{align}
  \cdots \to \pi_k(H) \to \pi_k(G) \to \pi_k(G/H) \to \pi_{k-1}(H) \to \pi_{k-1}(G) \to \cdots
\end{align}
From this exact sequence, we see that $\pi_k(G/H)=0$ for $k \leq d$ if $\pi_k(H) \to \pi_k(G)$ is an isomorphism for $k \leq d-1$ and a surjection for $k=d$. 

This is indeed the case for $d=4$ and $H=\mathrm{SU}(2)$ with the embedding $\mathrm{SU}(2) \subset G$ given by a long simple root. By our assumption, $\pi_0(G)=\pi_1(G)=0$. In general, $\pi_2(G)=0$. The $\pi_3(G) =\bZ$ classifies instantons of $G$, and any instanton can be deformed into $\SU(2)$ by Bott theorem and hence $\pi_3(G) =\pi_3(\mathrm{SU}(2))$. Also, we have $\pi_4(G)=0$ unless $G=\Sp(N)$, in which case $\pi_4(\Sp(N)) = \bZ_2$. This $\bZ_2$ is relevant for Witten $\SU(2)$ anomaly, and is again realized inside $\SU(2)=\Sp(1)$. 

We conclude that $G$ can be reduced to $\SU(2)$. Notice that in the above obstruction-theoretic argument, the CP or orientability plays almost no role because the analysis of each simplex is local, and each simplex is orientable (although not naturally oriented).\footnote{Implicitly, before gluing $\partial \sigma$ to $Y^{(k)}$, we need to perform gauge transformations by $\Pin^+(d+1)$ in the region of $Y^{(k)}$ where $\partial \sigma$ is glued, so that the transition function does not involve orientation-reversal elements of $\Pin^+(d+1)$.}

After reducing to the case of $\SU(2)$, the rest of the argument is the same as before. Our result may be summarized by the equality
\begin{align}
  \Omega_5^{\mathrm{Pin}^+ \ltimes G}(\mathrm{pt}) \simeq \Omega_5^{\mathrm{Spin}}(BG) \qquad (\pi_0(G)=\pi_1(G)=0).
\end{align}
More explicitly, the bordism group is $\mathbb{Z}_2$ for $G=\mathrm{Sp}(N)$, and otherwise it is $0$.

\section{String theory realization}\label{sec:string}
In this section, we discuss a string theory realization of the CP symmetry, or more precisely $\Pin^{+}(d) \ltimes G$ symmetry, for the case $d=4$, $G=\mathrm{Spin}(10)$, and chiral fermions in the representation $\mathbf{16}$. This is one of the grand unifield gauge group and its matter content. The purpose is to illustrate the existence of the CP symmetry in some compactifications of the $E_8 \times E_8$ heterotic string theory. If the total theory is anomaly-free~\cite{Witten:1985bt}, the four-dimensional theory after a topologically allowed compactification should also be anomaly-free. (See \cite{Tachikawa:2021mby,Yonekura:2022reu} for more discussions on these issues). Throughout this section, we assume familiarity with basic aspects of heterotic string theory compactifications (see e.g., \cite{Green:1987mn}). 

\subsection{Basics of compactification}
In the following discussions, we will sometimes use Lie algebra notations for symmetries instead of Lie group notations. For instance, we sometimes use $\mathfrak{so}(D)$ instead of $\mathrm{SO}(D), \mathrm{Spin}(D)$, or $\mathrm{Pin}^\pm(D)$ when we want to leave open the possibility of different global structures of the group. 

The CP symmetry in the case of Calabi-Yau compactifications has been already briefly discussed in e.g. \cite{Strominger:1985it,Dine:1992ya,Choi:1992xp,McNamara:2022lrw}, and here we consider slightly more general compactifications including non-Calabi-Yau cases. Anomalies are determined by topological properties rather than actual dynamics, so we will not care about equations of motion.\footnote{Calabi-Yau compactifications satisfy equations of motion with a flat $\mathbb{R}^4$, but more general compactifications may not. However, we may use such configurations as ininial conditions for time-dependent solutions, so they can be physically relevant. Also, when we consider path integrals, we need to sum over all field configurations including those which do not satisfy equations of motion.}

We consider a compactification of the ten-dimensional theory on a six-dimensional manifold $M$ to obtain a four-dimensional theory. 
The 3-form field strength $H$ of the $B$-field satisfies the Bianchi identity
\begin{align}
  \d H \propto \tr R \wedge R - \tr F \wedge F,\label{eq:bianchi}
\end{align}
where $R$ is the Riemann curvature 2-form, $F$ is the gauge field strength 2-form, and the trace is taken with an appropriate normalization. (This equation requires further refinement; see \cite{Witten:1985mj,Witten:1999eg,Yonekura:2022reu}.) This Bianchi identity implies that $F$ cannot be trivial when $R$ is nontrivial. 

Let us focus on one of the two $E_8$ gauge groups, whose Lie algebra is denoted by $\mathfrak{e}_8$. It contains a subalgebra $\mathfrak{so}(16) \subset \mathfrak{e}_8$ under which the adjoint representation is decomposed as
\begin{align}
  {\rm adj}(\mathfrak{e}_8) \to {\rm adj}(\mathfrak{so}(16)) \oplus \mathbf{128},
\end{align}
where $\mathbf{128}$ is one of the spinor representations of $\mathfrak{so}(16)$. The $\mathfrak{so}(16)$ contains a subalgebra $\mathfrak{so}(6) \oplus \mathfrak{so}(10)$ under which 
\begin{align}
  \mathbf{128} \to (\mathbf{4} \otimes \mathbf{16}) \oplus (\overline{\mathbf{4}} \otimes \overline{\mathbf{16}}). \label{eq:128decomp}
\end{align}
If the field strength $F$ takes value in the subalgebra $\mathfrak{so}(6) \subset \mathfrak{so}(16) \subset \mathfrak{e}_8$, the trace $\tr F \wedge F$ (as well as $\tr R \wedge R$) is taken in the defining representation of $\mathfrak{so}(6)$. We introduce a gauge field for $\mathfrak{so}(6)$ such that its bundle and connection is the same as those for the tangent bundle of $M$. Then the right-hand side of \eqref{eq:bianchi} vanishes, and we can take $H=0$. This is the standard embedding in heterotic string compactifications. 

The gaugino in ten dimensions is a Majorana-Weyl fermion in the adjoint representation of the gauge group. For our purposes, the important part of the representation is given by \eqref{eq:128decomp}.
The representation $\mathbf{4}$ is one of the spinor representations of $\mathfrak{so}(6)$, which is also the fundamental representation of $ \mathfrak{su}(4) = \mathfrak{so}(6)$.

Zero modes on $M$ of the gaugino give rise to chiral fermions in four dimensions. The gaugino is coupled to the $\mathfrak{so}(6)$ gauge bundle on $M$. The Atiyah-Singer index theorem implies, after some standard computations, that the number of (say) left-handed Weyl fermions in four dimensions in the representation $\mathbf{16}$ minus that in the representation $\overline{\mathbf{16}}$ is given by $\frac{1}{2} \chi(M)$, where $\chi (M)$ is the Euler number of the six-dimensional manifold $M$.\footnote{
  Let $\pm x_i~(i=1,2,3)$ be the Chern roots of the tangent bundle of $M$. Then, the $\hat{A}$-genus is given by $\hat{A}(R) = \prod_{i=1}^3 \frac{x_i/2}{\sinh(x_i/2)}$. The Chern character in the representation $\mathbf{4}$ is given by $\mathrm{ch}_{\mathbf{4}}(F) = e^{(x_1 + x_2 + x_3)/2} + e^{(x_1 - x_2 - x_3)/2} + e^{(-x_1 + x_2 - x_3)/2} + e^{(-x_1 - x_2 + x_3)/2}$. The index theorem states that the number of positive chirality zero modes of ${\bf 4}$ minus that of negative chirality zero modes is given by $\int_M \mathrm{ch}_{\mathbf{4}}(F) \hat{A}(R)$. By expanding this expression, we get $\frac{1}{2} \int_M  x_1 x_2 x_3 $, which is one-half of the Euler number $\chi(M)$ of $M$. The number of negative chirality zero modes of ${\bf 4}$ in six dimensions is equal to that of positive chirality zero modes of $\overline{\bf 4}$, and hence the difference between the number of positive chirality zero modes of ${\bf 4}$ and that of $\overline{\bf 4}$ is equal to $\frac12 \chi(M)$.
  }

We have obtained the net number of $\frac12 \chi(M)$ chiral fermions in four dimensions in the representation $\mathbf{16}$, modulo non-chiral fermions which are irrelevant for anomalies. This number can be unity, $\frac12 \chi(M)=1$. For instance, $M=S^6$ gives $\chi(S^6)=2$.\footnote{
The manifold $S^6$ is not Ricci-flat. As we mentioned above, we do not care about equations of motion in this section. However, it is possible to obtain a solution of equations of motion by introducing linear dilaton background. Similar configurations are discussed in a different context in \cite{Kaidi:2023tqo,Kaidi:2024cbx,Fukuda:2024pvu,Chikazawa:2025cis}.} 
Therefore, we see that the chiral fermion in four dimensions in the representation $\mathbf{16}$ can be realized from the heterotic string theory compactified on $M$.

\subsection{CP symmetry}
Now let us consider the CP symmetry. In ten dimensions, there is no CP symmetry since the ten-dimensional theory is chiral. However, the four-dimensional CP symmetry may be realized as a combination of a spatial reflection on $M$ and a reflection in the four-dimensional spacetime such that the total transformation is just a part of the ordinary Lorentz symmetry without orientation reversal in ten dimensions. For this purpose, there must be an isomorphism of $M$
\begin{align}
    \sigma_M : M \to M,
\end{align}
 that reverses the orientation of $M$. To be concrete, we consider $\sigma_M$ such that $(\sigma_M)^2=1$. For instance, $M=S^6$ has such an isomorphism given by $\vec{n} \to - \vec{n}$ where $\vec{n} \in S^6 \subset {\mathbb R}^7$ is the unit vector regarded as a coordinate system on $S^6$. 

We have taken the $\mathfrak{so}(6)$ gauge bundle to be the same as the tangent bundle of $M$. Therefore, the action of $\sigma_M$ on the tangent bundle induces the corresponding action on the $\mathfrak{so}(6)$ gauge bundle. A reflection is not an element of $\mathrm{SO}(6)$ or its spin lift $\mathrm{Spin}(6)$, and hence we must combine it with a reflection in (a group corresponding to) $\mathfrak{so}(10)$ to define the appropriate element in the total gauge group $\mathrm{Spin}(16) \subset E_8$. The reflection in $\mathfrak{so}(10)$ will play the role of an outer automorphism of $\mathfrak{so}(10)$, introduced in the general context in \eqref{eq:automorphism}. 

To distinguish spacetime Lorentz symmetry and gauge symmetry, we use subscripts ${\rm s}$ and ${\rm g}$ such as $\mathfrak{so}(D)_{\rm s}$ and $\mathfrak{so}(D)_{\rm g}$ for the former and the latter, respectively. Then we have 
\begin{align}
  [\mathfrak{so}(6)_{\rm s} \times \mathfrak{so}(4)_{\rm s}] \times [\mathfrak{so}(6)_{\rm g} \times \mathfrak{so}(10)_{\rm g}] \subset \mathfrak{so}(10)_{\rm s} \times \mathfrak{so}(16)_{\rm g}.
\end{align}
Suppose that we want to use an element $\mathsf{S} $ of (a group corresponding to) $\mathfrak{so}(6)_s$ for the definition of the isomorphism $\sigma_M : M \to M $.\footnote{
  This is only a simplification of what we actually need. The more precise situation is as follows. The isomorphism $\sigma_M$ induces an action on the tangent bundle of $M$, $(\sigma_M)_* : TM \to TM$, which further induces an action on the frame bundle associated to the tangent bundle.  This frame bundle must be lifted to the principal bundle associated to the spin structure of $M$, which we denote by $SM$. Then we need a lift $\mathsf{S}_* : SM \to SM$ of $(\sigma_M)_*$. This $\mathsf{S}_*$ is the one which we actually need for the definition of the CP symmetry in four dimensions. In the discussions of the main text, we pretend as if it is just given by an element $\mathsf{S}$ of (a group corresponding to) $\mathfrak{so}(6)_s$ for simplicity. The property $(\sigma_M)^2=1$ implies that $\mathsf{S}_*^2 = \pm 1$ depending on the topology of $M$ and the spin structure. However, as we will see below, the final conclusion does not depend on this sign.\label{ftn:lift}
}
For instance, if $\mathsf{S}$ is a relfection along a direction $\vec{m}=(m^a) \in \mathbb{R}^6$, then we explicitly take $\mathsf{S} ={m}_a \gamma^a$, where $\gamma^a$ are gamma matrices for $\mathfrak{so}(6)$. Then, we take the same element $\mathsf{S}$ also for $\mathfrak{so}(6)_{\rm g}$ because the tangent bundle and the gauge $\mathfrak{so}(6)_{\rm g}$ bundle are identified. Let $\mathsf{T}_{\vec n} = n_\mu \gamma^\mu$ be the reflection along a direction $\vec n \in \mathbb{R}^4 $ in four dimensions. Also, let $\mathsf{C}$ be a product of gamma matrices of $\mathfrak{so}(10)_{\rm g}$ such that for gamma matrices $\Gamma_I~(I=1,\ldots,10)$ of $\mathfrak{so}(10)_{\rm g}$, we have
\begin{align}
  \mathsf{C} \Gamma_I \mathsf{C}^{-1} = \Gamma_I^* \quad (I=1,\ldots,10) . \label{eq:Cdef}
\end{align}
More explicitly, if we take a representation of the Clifford algebra such that $\Gamma_I^*=\Gamma_I$ for odd $I$ and $\Gamma_I^*=-\Gamma_I$ for even $I$, then we may take $\mathsf{C} = \Gamma_1 \Gamma_3 \Gamma_5 \Gamma_7 \Gamma_{9}$ (see e.g. \cite{Weinberg:2000cr} for a textbook account of gamma matrices in general dimensions).\footnote{Another possible choice is to take $\mathsf{C} =\Gamma_2 \Gamma_4 \Gamma_6 \Gamma_8 \Gamma_{10}$, in which case $\mathsf{C} \Gamma_I \mathsf{C}^{-1} = - \Gamma_I^*$. The different between these two choices is just a center element of $\mathrm{Spin}(10)$, and is not important for the following discussions.}
Then we take the total transformation to be 
\begin{align}
  \sR_{\vec n} = (\mathsf{S}_{\rm s} \mathsf{T}_{\vec n} )(  \mathsf{S}_{\rm g} \mathsf{C} ), 
\end{align}
where the subscripts ${\rm s}$ and ${\rm g}$ for $\mathsf{S}$ indicate that they are elements of the Lorentz symmetry and the gauge symmetry, respectively. One can see by using exlicit gamma matrix representations that 
\begin{align}
  \mathsf{S}_{\rm s} \mathsf{T}_{\vec n} \in \mathrm{Spin}(10)_{\rm s}, \qquad \mathsf{S}_{\rm g} \mathsf{C} \in \mathrm{Spin}(16)_{\rm g}.
\end{align}
Therefore, the total transformation $\sR_{\vec n}$ is a ligitimate symmetry transformation in ten dimensions.

Recall that we are interested in the fermions in the representation \eqref{eq:128decomp}. Including the Lorentz symmetry, these fermions are in the bi-spinor representation of $\mathfrak{so}(10)_{\rm s} \times \mathfrak{so}(16)_{\rm g}$. In this case, one can see from the properties of gamma matrices that $\mathsf{S}_{\rm g} \mathsf{C} = - \mathsf{C} \mathsf{S}_{\rm g}$, $\mathsf{C}^2=1$, $\mathsf{S}_{\rm s} \mathsf{T}_{\vec n} = -\mathsf{T}_{\vec n} \mathsf{S}_{\rm s}$, and $\mathsf{S}_{\rm g}^2 = \mathsf{S}_{\rm s}^2 =\pm 1$ (see footnote~\ref{ftn:lift} for the reason why $\mathsf{S}^2= \pm 1$). Therefore, we obtain
\begin{align}
  \sR_{\vec n}^2 =\mathsf{T}_{\vec n}^2=(n_\mu \gamma^\mu)^2=1.
\end{align}
Therefore, we have confirmed that the relevant spacetime symmetry in four dimensions is $\mathrm{Pin}^+(4)$ rather than $\mathrm{Pin}^-(4)$. Moreover, the conjugation by $\mathsf{C}$ is an outer automorphism of $\mathfrak{so}(10)_{\rm g}$ which exchanges the two complex spinor representations of $\mathfrak{so}(10)_{\rm g}$ because of the relation \eqref{eq:Cdef}. The transformation $\mathsf{R}_{\vec n}$ contains $\mathsf{C}$, and hence the total symmetry group is
\begin{align}
  \Pin^+(4) \ltimes {\rm Spin}(10)
\end{align}
modulo one $E_8$ factor which we  have neglected. 

\section*{Acknowledgements}
The author would like to thank T.~T.~Yanagida for valuable discussions which have led to this work. 
 The work of KY is supported in part by JST FOREST Program (Grant Number JPMJFR2030, Japan), MEXT-JSPS Grant-in-Aid for Transformative Research Areas (A) ``Extreme Universe'' (No. 21H05188),and JSPS KAKENHI (21K03546).

\bibliographystyle{ytphys}
\def\url#1{\href{#1}{#1}}
%\baselineskip=.95\baselineskip
\bibliography{ref}

\end{document}